# On the Design and Analysis of Quaternary Serial and Parallel Adders

Anindya Das[1], Ifat Jahangir[2]
Department of Electrical and Electronic Engineering
Bangladesh University of Engineering and Technology
Dhaka, Bangladesh
anindya_088@yahoo.com[1], ifat00@gmail.com[2]

Masud Hasan
Department of Computer Science and Engineering
Bangladesh University of Engineering and Technology
Dhaka, Bangladesh
masudhasan@cse.buet.ac.bd

*Abstract*— **Optimization techniques for decreasing the time and area of adder circuits have been extensively studied for years mostly in binary logic system. In this paper, we provide the necessary equations required to design a full adder in quaternary logic system. We develop the equations for single-stage parallel adder which works as a carry look-ahead adder. We also provide the design of a logarithmic stage parallel adder which can compute the carries within $\log_2(n)$ time delay for *n* qudits. At last, we compare the designs and finally propose a hybrid adder which combines the advantages of serial and parallel adder.**

## I. INTRODUCTION

Among the various building blocks used in a microprocessor, adder is undoubtedly the most important one. Therefore performance enhancement of a microprocessor depends largely on the development of adder circuits. Since 1973, researchers [1]-[3] have been working relentlessly to optimize time delay, wiring complexity, fan-in, fan-out and chip area of the adder circuit. They used various mathematical techniques [2] and VLSI design optimizations like dynamic programming to minimize latency and area of the adder [3]. The most common technique of reducing the time delay is expressing carry of each bit as a function of propagate and generate terms of corresponding bit and its earlier bits. Almost all of these works are based on binary logic, so with the advent of multi-valued logic systems like ternary or quaternary logic, improved and efficient adders for these systems have become a prime necessity.

Multi-valued logic is an apparent extension of binary logic where any proposition can have more than 2 values. *Boolean algebra* in binary logic system is established by defining some fundamental operators which are fully analogous to union, intersection and compliment operations in set theory, but it is not readily expandable for multi-valued logic. In fact, a system with radix $2^k$ ($k \geq 2$) must have a lot of similarities with binary system [4], so a logic scheme having fundamental blocks that function similarly to their binary counterparts is surely advantageous for such logic systems.

In section III we have reviewed a new scheme for *quaternary logic* system which is closely related to binary system and was first proposed in [5]-[6]. Later we have utilized the operators and the associated algebra to design different kinds of adders. We have presented the expressions for half adder and *full adder* in quaternary logic system in section IV. The expressions for *propagate* and *generate* values of each *qudit* are also presented and a complete design of *single-stage carry look-ahead adder* in quaternary logic is demonstrated. Then we have proposed a design of *logarithmic stage carry-tree adder* which has time delay of $\log_2(n)$ due to its tree structure and optimal fan-in. In section V we have discussed the comparative performance of our proposed adders thoroughly with respect to different significant criteria at the following section. Finally, in section VI we have proposed two of many possible optimized designs: a *sparsity-4 adder* which is a combination of logarithmic stage carry-tree adder and single-stage carry look-ahead adder, and a *hybrid adder* having both serial and parallel carry generation techniques combined in one adder.

## II. EARLIER WORKS ON QUATERNARY ADDER

Although the concept of multi-valued logic has been existent for a long time, it has introduced some challenges that prevented the vast developments of multi-valued adders, specially quaternary fast adders, which is extremely rare in the literature. For quaternary fast/parallel adders, the main limitation is the lack of physically implementable circuit models and the inherent complexity associated with carry-look-ahead/carry-skip trees. Yet several works on quaternary adders were published during the last few decades, those adders were built on different technologies- including all-optical, *CMOS current-mode* and *voltage-mode* circuits and *quantum gates*.

In a work by Awwal *et al.* [7], a high speed parallel adder was presented that could perform *carry-free addition* of two modified signed digit quaternary numbers. In that work, as a mean of physical implementation, binary coupled quaternary optical system was used with experimental results obtained through an *optical PLA*. Current [8] proposed a circuit that realized the current-mode quaternary full adder function with transparent latching in a standard polysilicon-gate CMOS technology. Another work by Silva *et al.* [9] demonstrated a voltage-mode quaternary full adder circuit using three power supply lines and *multi-threshold transistors*. Wu, Chen and Prosser [10] investigated into quaternary CMOS full adder based on transmission function theory, where instead of conventional CMOS switching operation, they utilized *transmission gates* to control signal flow.

As a promising and evolving technology, many researchers have chosen quantum gates for designing and implementing quaternary adders. Khan [11] proposed a recursive method of hand synthesis of *reversible quaternary full-adder circuit*. In his novel work he proposed that macro-level quaternary controlled gates built on the top of ion-trap realizable 1-qudit quantum gates and 2-qudit *Muthukrishnan–Stroud* gates can be used for implementing full adders and parallel adders.

## III. QUATERNARY ALGEBRA

In this section we have reviewed a novel quaternary logic that was proposed in [5], [6]. Quaternary states (0, 1, 2, 3) can be imagined as 2-bit binary equivalents 00, 01, 10, 11. If the bits of the binary equivalent interchange their positions and still the quaternary state remains unchanged, then it is said to have *binary symmetry*; otherwise it is *asymmetrical*. Thus 0, 3 are symmetrical and 1, 2 are asymmetrical. A single quaternary digit is called a qudit when it is expressed as a number.

The algebra we have developed so far in [5] has two types of operators – basic operators and special operators.

The basic quaternary operators are very similar to binary operators and they are obtained from Boolean algebra. They operate as *bitwise binary operators* working on 2-bit operands, as shown in Table I. The basic operators (Fig. 1) are *OR*, *AND*, *BASIC INVERTER* and *XOR*. Their derivatives are *BASIC NOR*, *BASIC NAND* and *BASIC XNOR*. The inverter is named as basic inverter since there are other special inverters or *compound inverters* in quaternary system to facilitate complex logic design.

The special operators (Fig. 2) are all unary operators. They are -

(a) *Outward Inverter* or *Full Inverter*
(b) *Inward Inverter* or *Half Inverter*
(c) *Binary Bitswap*

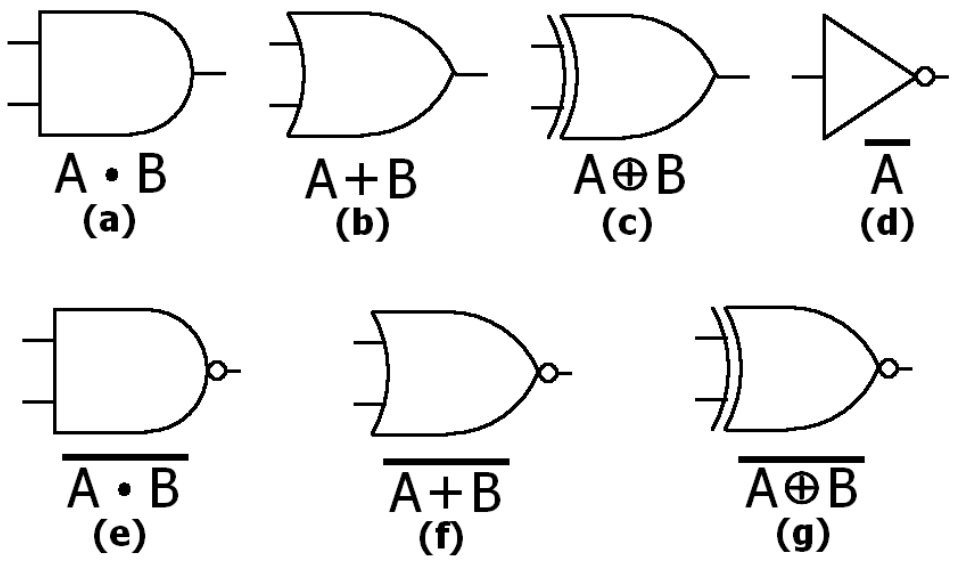

Figure 1. Circuit symbols for the basic quaternary operators. (a) AND, (b) OR, (c) XOR, (d) BASIC NOT, (e) BASIC NAND, (f) BASIC NOR, (g) BASIC XNOR .

The outward inverter inverts the input just like the basic inverter, but after that it changes the asymmetrical values to nearest symmetrical values.

The inward inverter inverts the input just like the basic inverter; but after that, unlike outward inverter, it changes the symmetrical values to nearest asymmetrical values.

The binary bitswap swaps the two bits of the binary equivalent of the quaternary operand. It leaves the symmetrical numbers unchanged but inverts (basic inversion) the asymmetrical numbers.

Like basic inverter, bitswap operator and special inverters can be cascaded with other operators to form compound operators such as inward NAND, outward NOR, bitswap XOR, etc. Circuit symbols for special operators and their compound derivatives are shown in Fig. 2.

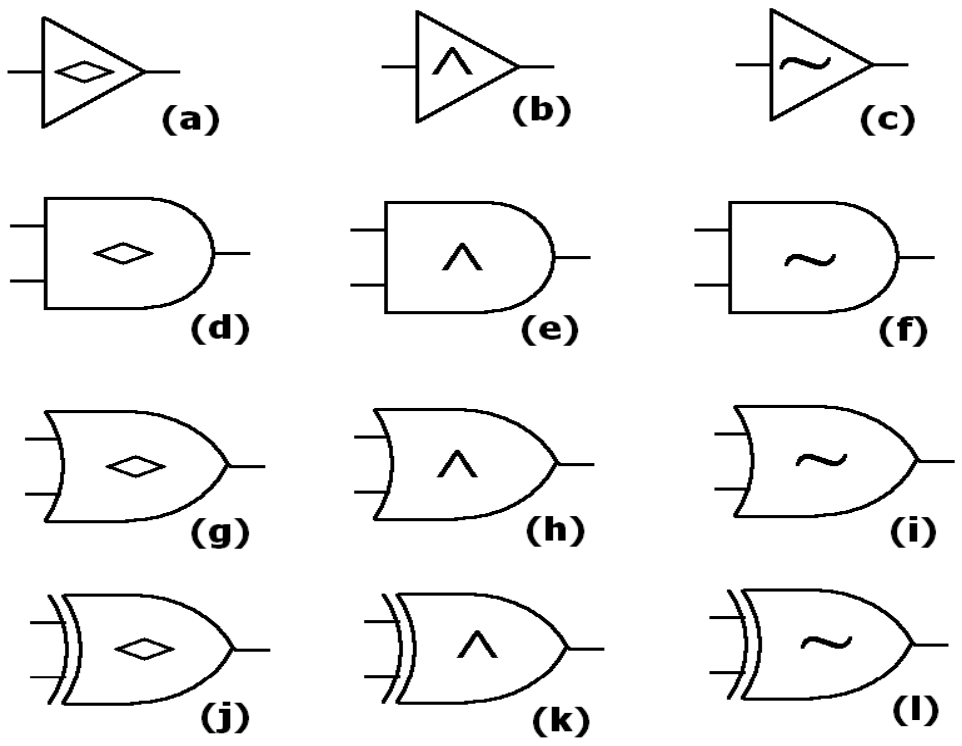

Figure 2. Circuit symbols for the special quaternary operators and their derivatives: (a) Inward or Half Inverter; (b) Outward or Full Inverter; (c) Binary Bitswap; (d) Inward NAND; (e) Outward NAND; (f) Bitswap AND; (g) Inward OR; (h) Outward OR; (i) Bitswap OR; (j) Inward XNOR; (k) Outward XNOR; (l) Bitswap XOR.

TABLE I. QUATERNARY MULTI-INPUT OPERATORS

| A | B | AND | OR | XOR | BASIC NAND | BASIC NOR | BASIC XNOR |
|---|---|---|---|---|---|---|---|
| 0 | 0 | 0 | 0 | 0 | 3 | 3 | 3 |
| 0 | 1 | 0 | 1 | 1 | 3 | 2 | 2 |
| 0 | 2 | 0 | 2 | 2 | 3 | 1 | 1 |
| 0 | 3 | 0 | 3 | 3 | 3 | 0 | 0 |
| 1 | 1 | 1 | 1 | 0 | 2 | 2 | 3 |
| 1 | 2 | 0 | 3 | 3 | 3 | 0 | 0 |
| 1 | 3 | 1 | 3 | 2 | 2 | 0 | 1 |
| 2 | 2 | 2 | 2 | 0 | 1 | 1 | 3 |
| 2 | 3 | 2 | 3 | 1 | 1 | 0 | 2 |
| 3 | 3 | 3 | 3 | 0 | 0 | 0 | 3 |

Fig. 3 illustrates the transfer characteristics of the special operators. Their mathematical definitions are given below:

$$\text{Inward Inverter,}\quad a' = \begin{cases} \overline{a}.2 & ; a < 2 \\ \overline{a} + 1 & ; a > 1 \end{cases} \tag{1}$$

$$\text{Outward Inverter,}\quad \hat{a} = \begin{cases} \overline{a} + 3 & ; a < 2 \\ \overline{a}.0 & ; a > 1 \end{cases} \tag{2}$$

$$\text{Binary Bitswap,}\quad \tilde{a} = \begin{cases} \overline{a} & ; \ a \ \text{asymmetric} \\ a & ; \ a \ \text{symmetric} \end{cases} \tag{3}$$

The special operators are necessary for designing important logic blocks like adder, *multiplexer*, *decoder*, etc [6]. Among them, binary bitswap is the most important one. It is used to define equality operator, which gives the sum of product (SOP) expression of a quaternary function as shown in [5]. The equality operator is defined as

$$a^b = b^a = \begin{cases} \bar{a} . a \ ; a \neq b \\ \bar{a} + a \ ; a = b \end{cases} = \begin{cases} 0 \ ; a \neq b \\ 3 \ ; a = b \end{cases} \qquad (4)$$

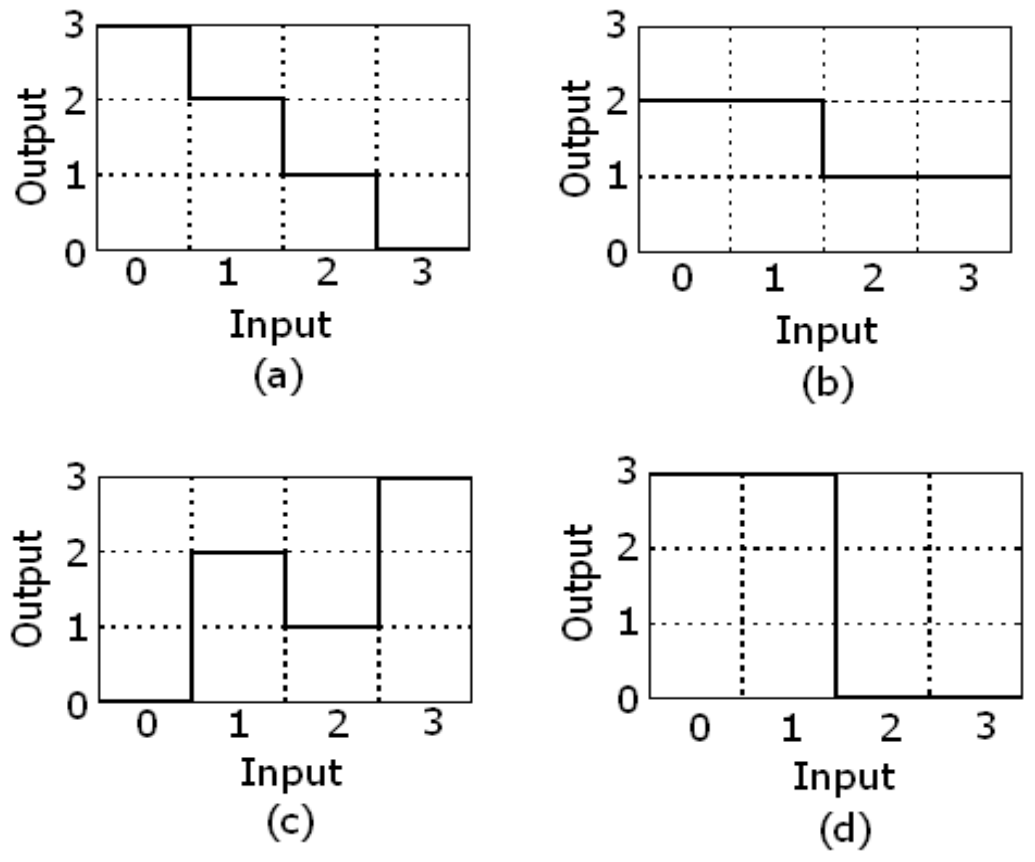

Figure 3. Transfer functions of the quaternary inverters and inverter-like operators: (a) Basic Inverter; (b) Inward or Half Inverter; (c) Binary Bitswap; (d) Outward or Full Inverter.

### *A. Expansion of De Morgan's Theorem*

De Morgan's theorem can be expanded in quaternary logic. Since basic inverter acts as bitwise binary operator, it obeys De Morgan's Theorem. This expansion is also true for outward inverter.

$$\widehat{(a + b)} = \hat{a} . \hat{b} \ \text{ and } \ \widehat{(a . b)} = \hat{a} + \hat{b} \qquad (5)$$

But it does not hold true for inward inverter and binary bitswap. Actually, the distribution of inward inverter over basic operators can not be expressed through any simple relation. Therefore

$$\left.\begin{array}{l} (a + b)' \neq a'.b' \ \text{ and } \ (a.b)' \neq a' + b' \\ (a + b)' \neq a' + b' \ \text{ and } \ (a.b)' \neq a'.b' \end{array}\right\} \qquad (6)$$

### *B. Interchangeability of Unary Operators*

These properties are useful when the special operators are cascaded with other special operators or basic inverter.

*1)* The order of basic inversion and any other special operator can be altered.

$$\hat{\bar{a}} = \bar{\hat{a}}, \ \tilde{\bar{a}} = \bar{\tilde{a}}, \ (\bar{a})' = (\overline{a'}) \qquad (7)$$

*2)* The order of any two special operators cannot be altered.

$$\tilde{\hat{a}} \neq \hat{\tilde{a}}, \ (\widetilde{a'}) \neq (\tilde{a})', \ (\widehat{a'}) \neq (\hat{a})' \qquad (8)$$

### *C. Distributive Nature of Bitswap Operator*

Bitswap operator distributes itself over any basic operator. This is very important property because bitswap is the most widely used special operator in our logic system. Its distributive nature makes many complex designs simple.

**I.** $\tilde{a} \oplus \tilde{b} = \widetilde{(a \oplus b)}$ (9)

**II.** $\widetilde{(a + b)} = \tilde{a} + \tilde{b}$ and $\widetilde{(a . b)} = \tilde{a} . \tilde{b}$ (10)

## IV. Quaternary Adders

Since the performance of adders play dominant role in the overall performance of many digital systems, special care has to be taken for proper optimization of adders. Speed, physical limitations in VLSI technology and the purpose of adders in a system are some key points that must be considered at the time of designing. A complete description of quaternary adders requires all possible situations to be considered and different adders are needed to be designed to handle different contingencies. For that reason we start our discussion with single qudit full adder and ripple carry adder. Then in subsection A we propose the design of single-stage parallel adder, and in subsection B the design of logarithmic stage carry-tree adder is introduced.

A full adder reduces to a half adder if carry input is set to zero. The first half of the following truth table (Table II) for full adder has zero carry input and hence represents a half adder.

TABLE II. Truth Table of Quaternary Full Adder

| **Carry-in = 0 (Half adder)** | | | | | | | |
|---|---|---|---|---|---|---|---|
| ***A*** | ***B*** | ***S*** | ***C*** | ***A*** | ***B*** | ***S*** | ***C*** |
| 0 | 0 | 0 | 0 | 1 | 2 | 3 | 0 |
| 0 | 1 | 1 | 0 | 1 | 3 | 0 | 1 |
| 0 | 2 | 2 | 0 | 2 | 2 | 0 | 1 |
| 0 | 3 | 3 | 0 | 2 | 3 | 1 | 1 |
| 1 | 1 | 2 | 0 | 3 | 3 | 2 | 1 |
| **Carry-in = 1** | | | | | | | |
| ***A*** | ***B*** | ***S*** | ***C*** | ***A*** | ***B*** | ***S*** | ***C*** |
| 0 | 0 | 1 | 0 | 1 | 2 | 0 | 1 |
| 0 | 1 | 2 | 0 | 1 | 3 | 1 | 1 |
| 0 | 2 | 3 | 0 | 2 | 2 | 1 | 1 |
| 0 | 3 | 1 | 1 | 2 | 3 | 2 | 1 |
| 1 | 1 | 3 | 0 | 3 | 3 | 3 | 1 |

We derived the expressions for half adder in [6], which results in the following equations after slight modification:

$$S = A \oplus B \oplus (\widetilde{A.B.1}) \qquad (11a)$$

$$C = \big( (A.B)' + A.B.(\widetilde{A \oplus B}) \big).1 \qquad (11b)$$

Although the basic operators of our proposed logic scheme work as bitwise binary operators, the expression of sum is not fully analogous to the expression of binary half adder. It happens because if we assume a qudit as a pair of bits, there will be an internal carry when each of the operands is either 1 or 3. So the expression of sum for quaternary half adder is not just the xor of two qudits as it is the case in binary logic.

It is intuitive to cascade two half adders of Fig. 4(a) to build a full adder but that will unnecessarily increase the number of gates. So we have derived the following expressions for full-adder that considers both operands and input carry simultaneously:

$$T = A \bullet B + B \bullet C_{in} + C_{in} \bullet A \quad (12a)$$

$$S = A \oplus B \oplus C_{in} \oplus (\widetilde{T \bullet 1}) \quad (12b)$$

$$C_{out} = ((A \bullet B)' + T \bullet (\widetilde{A \oplus B})) \bullet 1 \quad (12c)$$

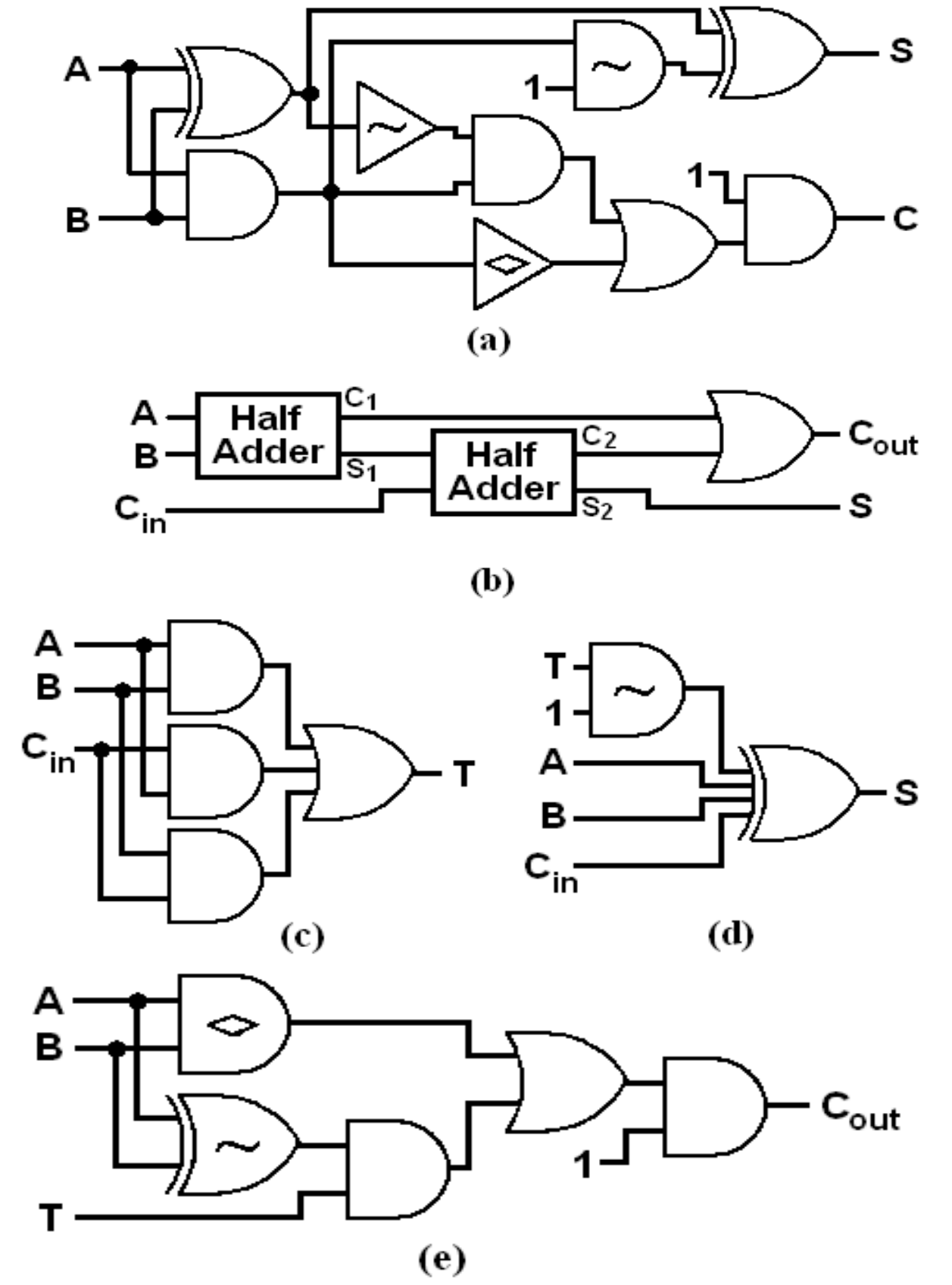

Figure 4. (a) Half-adder; (b) Cascade adder where two half-adders are cascaded to form a full-adder; (c) –(e) components of a full-adder in (12).

It is notable that the equations in (12b) are (12c) are expanded version of (11a) and (11b), taking carry-in into account. These equations are derived from the truth table in Table II, in the same way (11a) and (11b) are derived.

The cascaded design of full adder shown in Fig. 4(b) requires 19 gates, where the other design in Fig 4(c)-(e) takes 11 gates only. If we restrict fan-in to 2, even then the number of gates for the second design will be 14, which is still lower. The gate delay for the cascaded design is 9, but for the second design it is only 5 (or 6, if fan-in is restricted to 2). So this full adder is more efficient than two cascaded half adders in all respects. Hence it is used in *ripple carry adder* as shown in Fig. 5, where the carry generated at $i$-th qudit is the input carry for ($i$+1)-th qudit. So the ($i$+1)-th qudit of the summation cannot be evaluated until the carry-out from $i$-th qudit is generated. Hence an $n$-qudit adder will have a gate delay linearly increasing with the increase of qudits in the operands, which is a severe disadvantage of ripple carry adder.

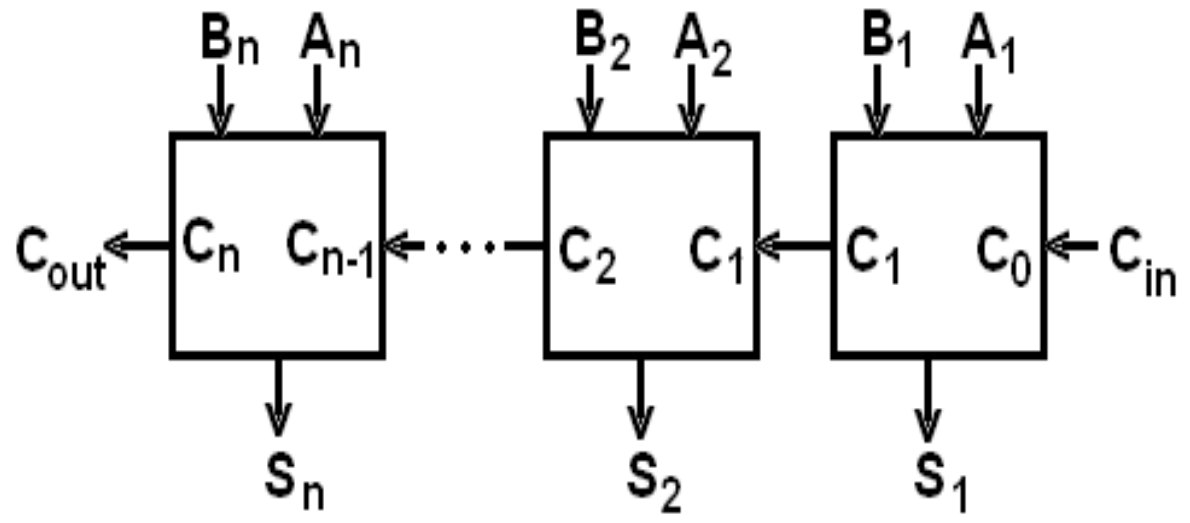

Figure 5. Ripple carry serial adder using full-adders.

To overcome the problem of gate delay in ripple carry adder, *parallel carry generators* are needed, which generates carries for all qudits fast enough so that we can generate all sum values simultaneously using (12a) and (12b). The gate delay should also be as small as possible. There are two key terms for representing the whole process:

***Generate*** *: It indicates that a carry is generated by adding i-th qudits of both operands regardless of the input carry at i-th position. This is represented by* $G_i$ *for i-th qudit.*

***Propagate*** *: It indicates that no carry is generated by adding i-th qudits of two operands, but an input carry at that position will be propagated to next bits. This is represented by* $P_i$ *for i-th qudit.*

It is obvious from the definition of generate that its expression will be the same as the expression of carry of half adder.

From the definition of propagate, we can conclude that a carry will propagate from i-th qudit to the next qudit only when $A_i$+$B_i$ will be equal to 3. So checking the value of XOR($A_i$,$B_i$) will serve our purpose in this case. Therefore we can use the following equation which is evident from the definition of bitswap operator:

$$\tilde{a} \,.\, a = \begin{cases} 3 \ ; \ a = 3 \\ 0 \ ; \ a \neq 3 \end{cases} \quad (13)$$

Thus the following expressions of $G_i$ and $P_i$ are derived in (14b) and (14c):

$$P_i^* = A_i \oplus B_i \quad (14a)$$

$$P_i = P_i^* \bullet \widetilde{P_i^*} \quad (14b)$$

$$G_i = ((A \bullet B)' + A \bullet B \bullet \widetilde{P_i^*}) \bullet 1 \quad (14c)$$

### A. Single-stage Parallel Adder:

In the fastest parallel adder (Fig. 6), the carry-out from $i$-th qudit is calculated using the following expression:

$$C_i = G_i + P_i \cdot C_{i-1} \tag{15}$$

After successive substitution and simplification, (14) can be generalized as

$$C_n = G_n + \sum_{i=1}^{n-1} G_i \cdot \left( \prod_{j=i+1}^{n} P_j \right) + \left( \prod_{j=1}^{n} P_j \right) \cdot C_0 \tag{16}$$

This is the simplest design for parallel carry generation, especially for small number of qudits. But if the adder becomes a little bit larger, fan-in increases linearly. Besides the *product terms* involving propagate expressions $P_j$ contain much redundancy with the increase of qudits. So any practical design using this adder will require small blocks of parallel adders, and carry ripples from lower order blocks to higher order blocks just like the ordinary ripple carry adder. The gate delay and fan-in is kept significantly low in this adder design.

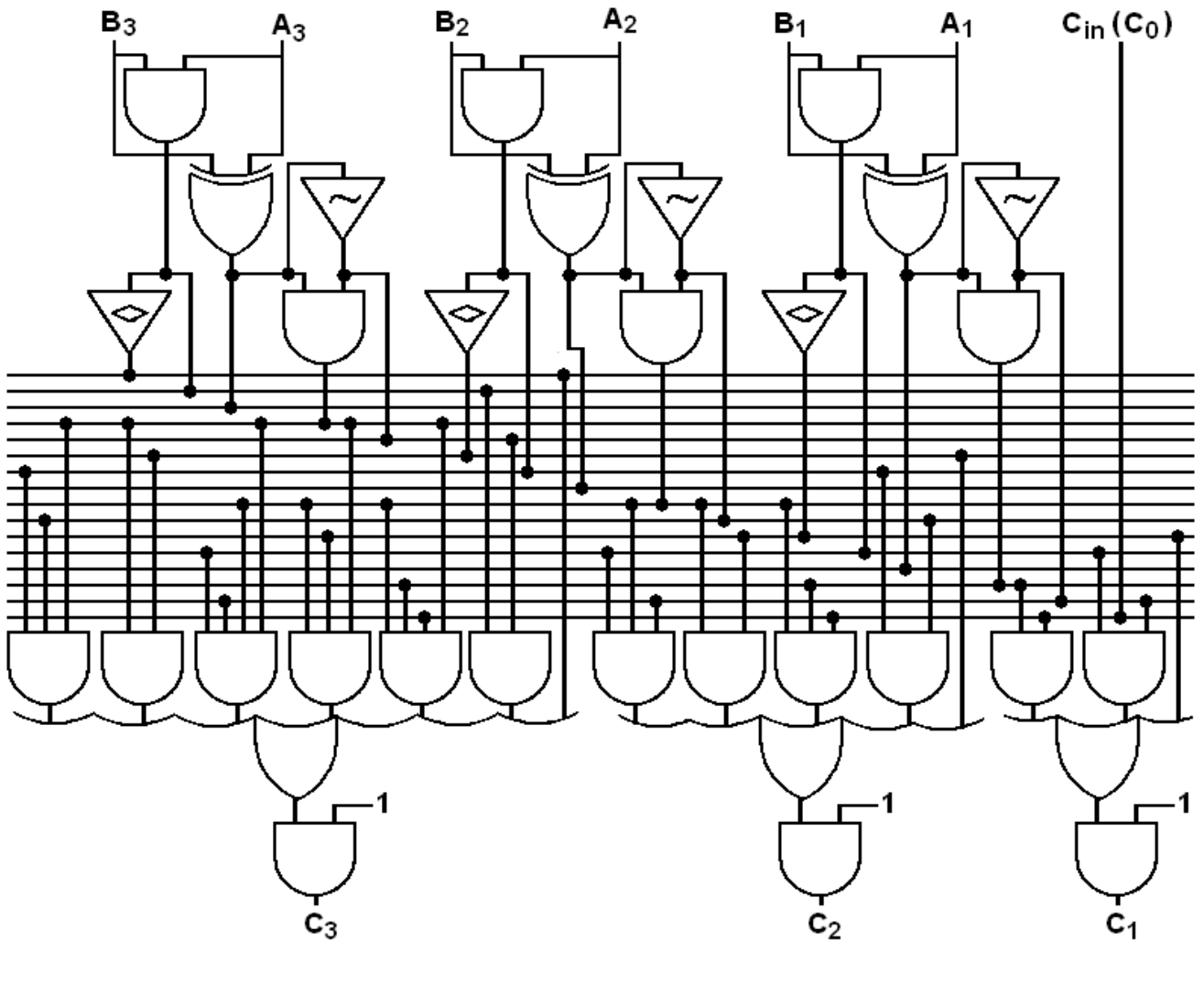

Figure 6. 3-qudit single-stage parallel adder.

### B. Logarithmic Stage Carry-tree Adder:

In order to overcome the problems of high fan-in and redundancy, the fast adder can be designed as a tree leaving the propagate and generate expressions unchanged. We have derived the recursive equations in (17b) and (17c) to represent the product tree and carry tree of this adder. These tree structures were first proposed by Kogge *et al.* [2].

$$m = 2^{\text{floor}(\log_2 |i-j|)} \tag{17a}$$

$$P(i,j) = \begin{cases} P(i,j\text{-}m) \cdot P(j\text{-}m+1,j), & i < j \\ P_i, & i = j \end{cases} \tag{17b}$$

$$Q(i,j) = \begin{cases} Q(i,i\text{-}m+1) + Q(i\text{-}m,j) \cdot P(i\text{-}m,i\text{-}1), & i > j \\ G_{i-1}, & i = j \end{cases} \tag{17c}$$

Each internal node of the carry tree ($Q(i,j)$ where $i \neq j$ ) has two children as shown in Fig. 7(a), where either the first parameter reduces by $2^{\text{floor}(\log2(i-j))}$ or the second parameter increases by the same amount and other parameter remains unchanged. The leaf nodes of the tree are $Q(i,i)$ which contain the generate values of each qudit except $Q(1,1)$ which is actually the carry-in. Thus $Q(i,1)$ contains the carry value for $i$-th qudit.

The term $P_i$ in (17b) denotes the propagate value of $i$-th qudit and $P(i,j)$ is the product term evaluated using a product tree shown in Fig. 7(b). The product tree is formed using (17b) in an almost similar manner where every leaf node $P(i,i)$ contains the propagate value $P_i$ for $i$-th qudit.

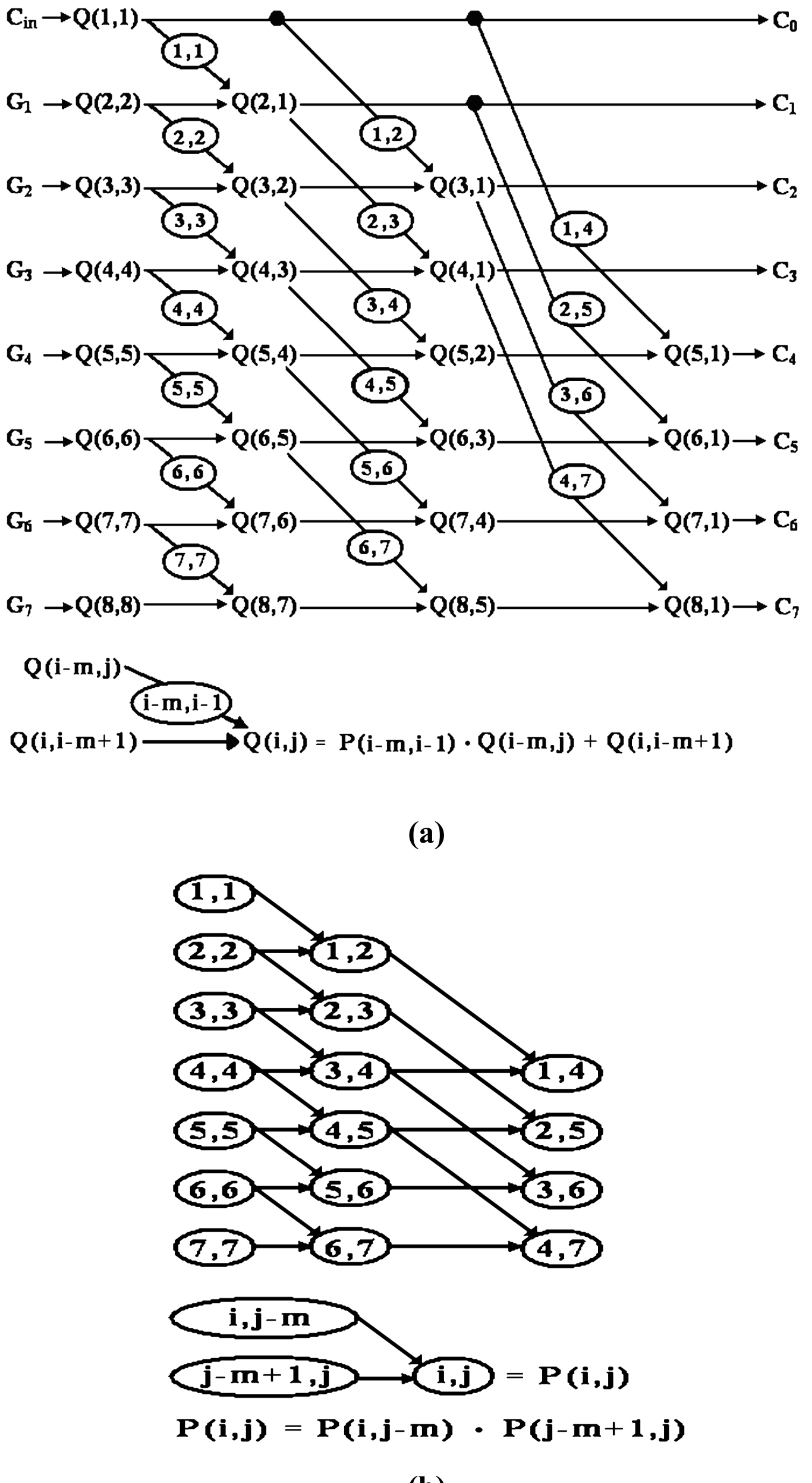

Figure 7. (a) 7-qudit logarithmic stage carry-tree adder; (b) product tree for this adder.

**Lemma 1:** It is possible to reach the leaf nodes ($Q(i,i)$) from any internal node $Q(i,j)$ within at most floor($\log_2(i\text{-}j)$)+1 steps.

**Proof:** Either $i$ is reduced or $j$ is increased by the largest power of 2 which is less than or equal to ($i$-$j$). So the difference between the two parameter decreases in every step by the largest power of 2. Any number $n$ can be represented using floor($\log_2 n$)+1 bits in binary representation. Decreasing $n$ by the largest power of 2 which is not greater than $n$ is nothing but erasing the left most bit of $n$. So if we erase the bits one by one from the left, we need floor($\log_2 n$)+1 steps to make $n$ zero. So within at most floor($\log_2(i\text{-}j)$)+1 steps, we can reach $Q(i,i)$ from $Q(i,j)$.

As the product tree is generated in an almost similar manner, this lemma will also be applicable for the product tree. So this adder can calculate $n$ carries within floor($\log_2 n$)+1 gate delays because $Q(i,j)$ converge to $Q(i,i)$ within floor($\log_2 n$)+1 depth of the tree.

## V. PERFORMANCE ANALYSIS:

The serial and parallel adders described so far differ widely from one another in performance and complexity. Usually efficient adders are more complex in nature, so there is always some trade off in speed, cost-efficiency or complexity.

Some important parameters in performance evaluation are gate delay, wiring and gate cost, largest fan-in in a circuit, etc. Here we develop the expressions for these quantities for each of the adders described in the preceding sections. Since the expression of sum is same for all the adders we are considering here, we only take the carry generation circuit into account.

Gate delay affects the performance of an adder very severely so it is of maximum priority to reduce the delay as much as possible. In our analysis we have assumed all gates to have same amount of delay. From Fig. 6, we can see that single-stage parallel adder has a fixed number of gate stages regardless of the number of qudits. The delay for this adder is 6 and it is constant for any qudit.

For serial ripple carry adder, every qudit incurs a delay of 5 and for $n$ qudits the delay is $5n$.

For logarithmic stage parallel adder, the delay at the propagate and generate stage is 4 and then the delay inside the adder tree is 2[ceil($\log_2 n$)]. Here the product tree is evaluated in parallel with the main adder tree, so it does not add any extra delay. The total delay for logarithmic adder is 4 + 2 [ceil($\log_2 n$)].

In Fig. 8, we have plotted the gate delay for each of these adders against the size of the adder. For small adders, the difference in delay between serial and parallel adders is not significant, but with the increase of size, parallel adders begin to dominate in speed. Single-stage parallel adder is faster than logarithmic stage parallel adder for large adders.

If ripple carry adder is used, then the number of gates used for generating each carry is 9 and number of inputs is 19.

For single-stage parallel adder the cost of carry generation is different for different qudits, and it increases rapidly. For $i$-th qudit, number of gates = $2i + 7$ and number of inputs = $i^2 + 5i + 11$.

So, for a $n$-qudit adder, total number of gates = $n^2 + 8n$ and total number of inputs = $n^3/3 + 3n^2 + 41n/3$.

For logarithmic stage carry-tree adder, the calculation of cost is more complex than other adders. For $n$ qudit adder, the initial stage of the tree has $n$ propagate terms and $n$ generate terms. Then at the first stage of product tree following the initial propagate stage, there are $n$-1 gates. If $s$ = floor ($\log_2 n$), there are $s$ stages after the propagate stage and the total number of gates in the product tree is $s(n+1) - 2^{s+1} + 2$. The number of inputs is just double i.e. $2s(n+1) - 2^{s+2} + 4$.

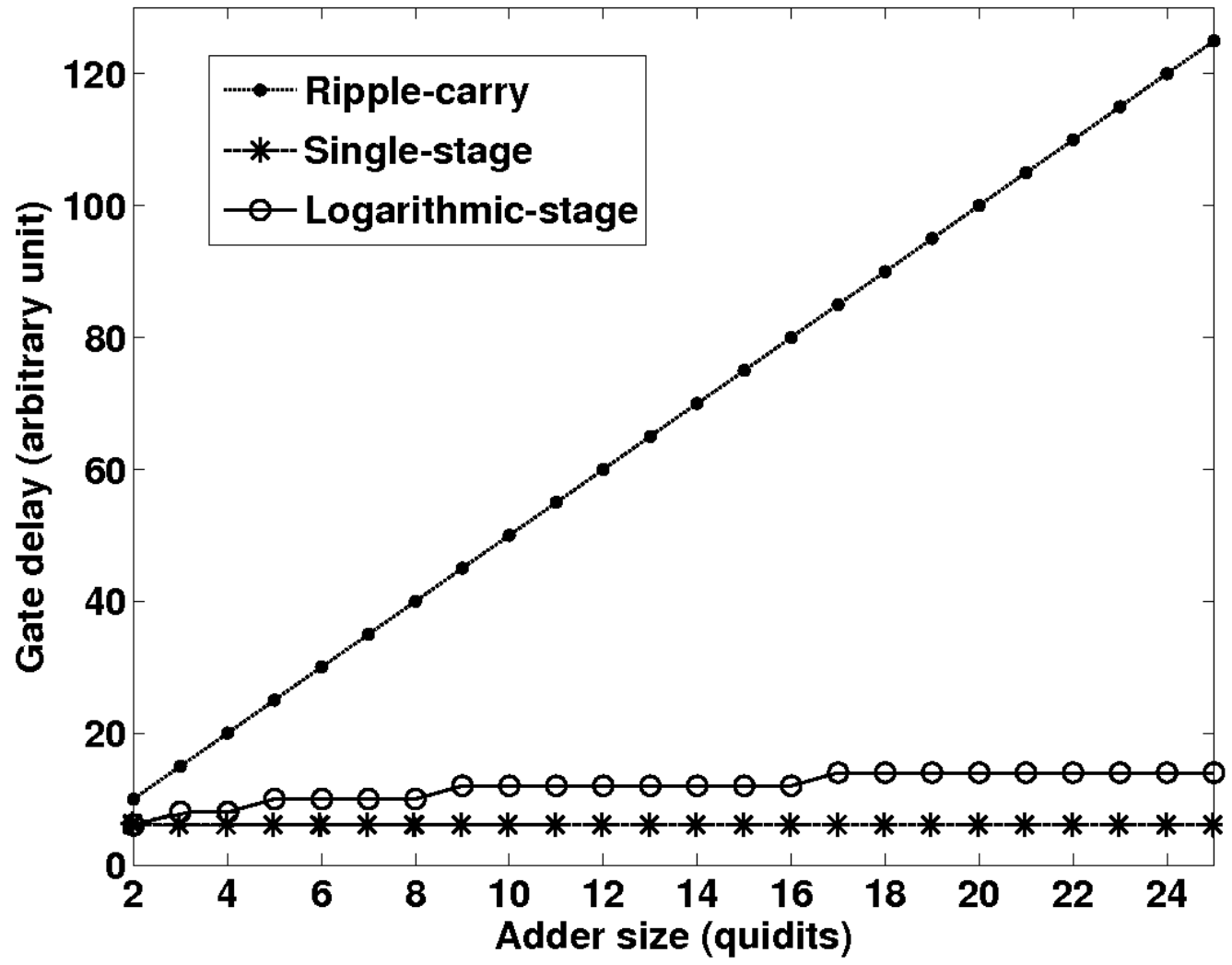

Figure 8. Comparison of gate delay for different adders.

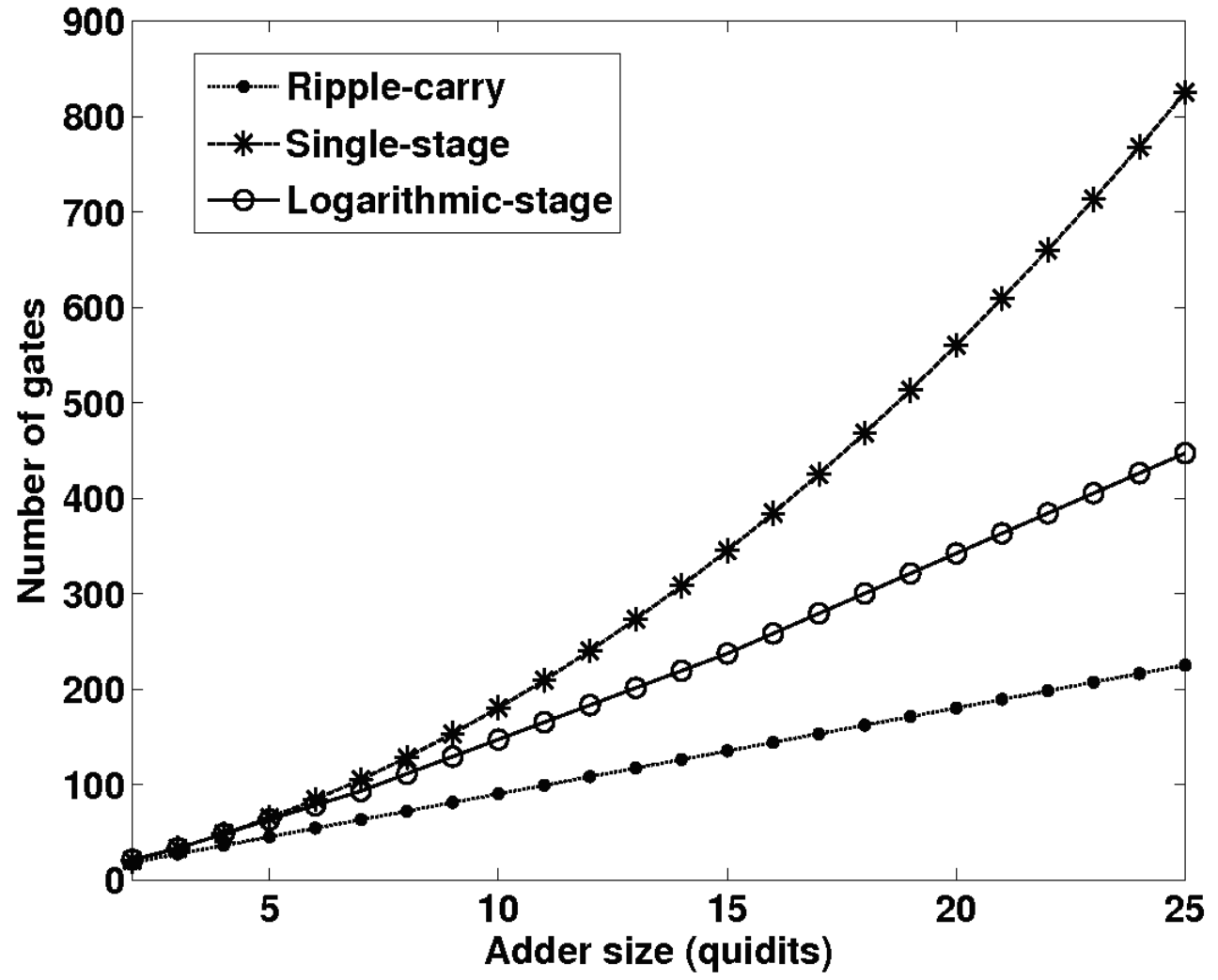

Figure 9. Comparison of gate cost for different adders.

Again the carry tree has generate blocks and carry-in at the initial stage. Each node from subsequent stages comprises of two gates connecting two nodes from its earlier stage. If $s$ = floor ($\log_2 n$), the total number of gates in the carry tree is $2(sn + s + n) - 2^{s+2} + 4$ and the number of inputs is $4(sn + s + n) - 2^{s+3} + 8$.

The generate and propagate blocks at the initial stages of logarithmic stage carry-tree adder have $7n$ gates and $16n$ inputs for $n$-qudits addition.

Hence the total numbers of gates and inputs for $n$-qudit adder are $6(1 - 2^s) + 3(sn + s + 3n)$ and $12(1 - 2^s) + 6(sn + s + 3n) + 2n$ respectively, with $s$ = floor ($\log_2 n$). Fig. 9 and Fig. 10 compare the number of gates and number of gate inputs respectively for all three adders.

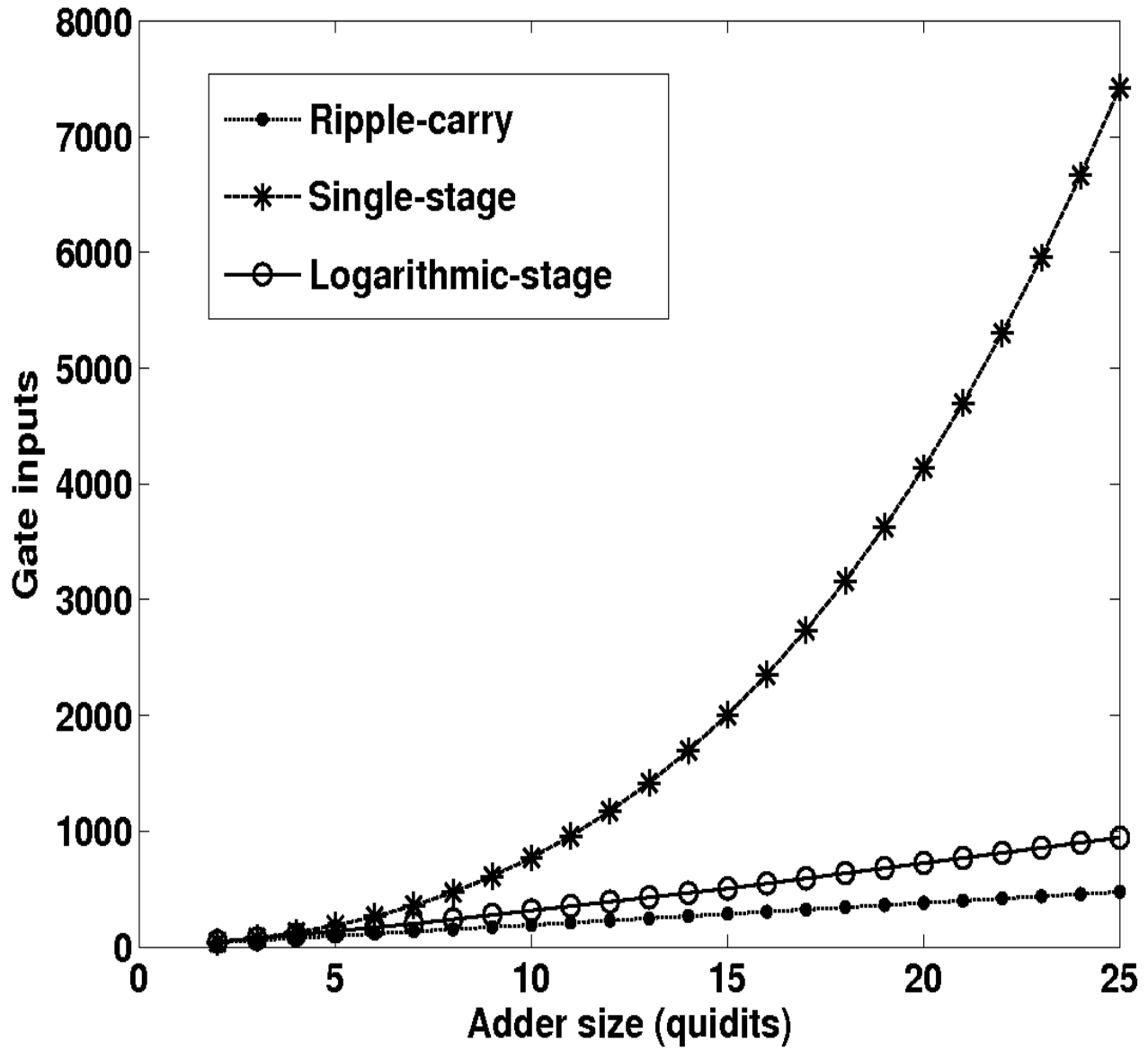

Figure 10. Comparison of wiring cost for different adders.

Largest fan-in for ripple carry adder is 3, considering only the carry generating circuit. This becomes $2n$+1 for $n$-qudit single-stage parallel adder. Again for logarithmic stage carry-tree adder, fan-in is limited to 3. So we see that adverse effects of fan-in are more pronounced in larger single-stage parallel adders.

## VI. Design Optimization of Large Adders:

In this section we propose some schemes to optimize the design of large adders. For example, we can achieve such optimization by modifying the carry tree of logarithmic stage carry-tree adder. Optimization of adder includes the increase of the radix and sparsity of the carry tree. Radix of the tree is defined as the maximum number of children of any node. Increment of the radix of the carry tree results in decrement of the number of stages but increase of fan-in and delay in each stage. So we fix the radix as 2.

Sparsity of an adder is defined as the number of every $n$-th carry generated by the carry tree. We can efficiently reduce root congestion and area by increasing the sparsity of the adder. We propose here a sparsity-4 adder which can be constructed as follows (Fig. 11, 12).

Fig. 11 demonstrates how it can calculate 3rd and 7th carry and Fig. 12 demonstrates the product tree of this carry tree. We can easily extend it to generate every ($4k$-1)-th carry where $k$ is a positive integer less than or equal to $n$/4 for any large $n$. Each carry generated by this sparsity-4 adder feeds the carry in of a single-stage carry look-ahead adder or a ripple carry adder to generate carry at every $4k$, $4k$+1 and $4k$+2 qudit positions.

Another optimization is possible using ripple carry adders. Here instead of rippling through every single qudit, carry ripples through blocks of multiple qudits and inside the block carry is generated by either of the parallel adders. This scheme is shown in Fig. 13.

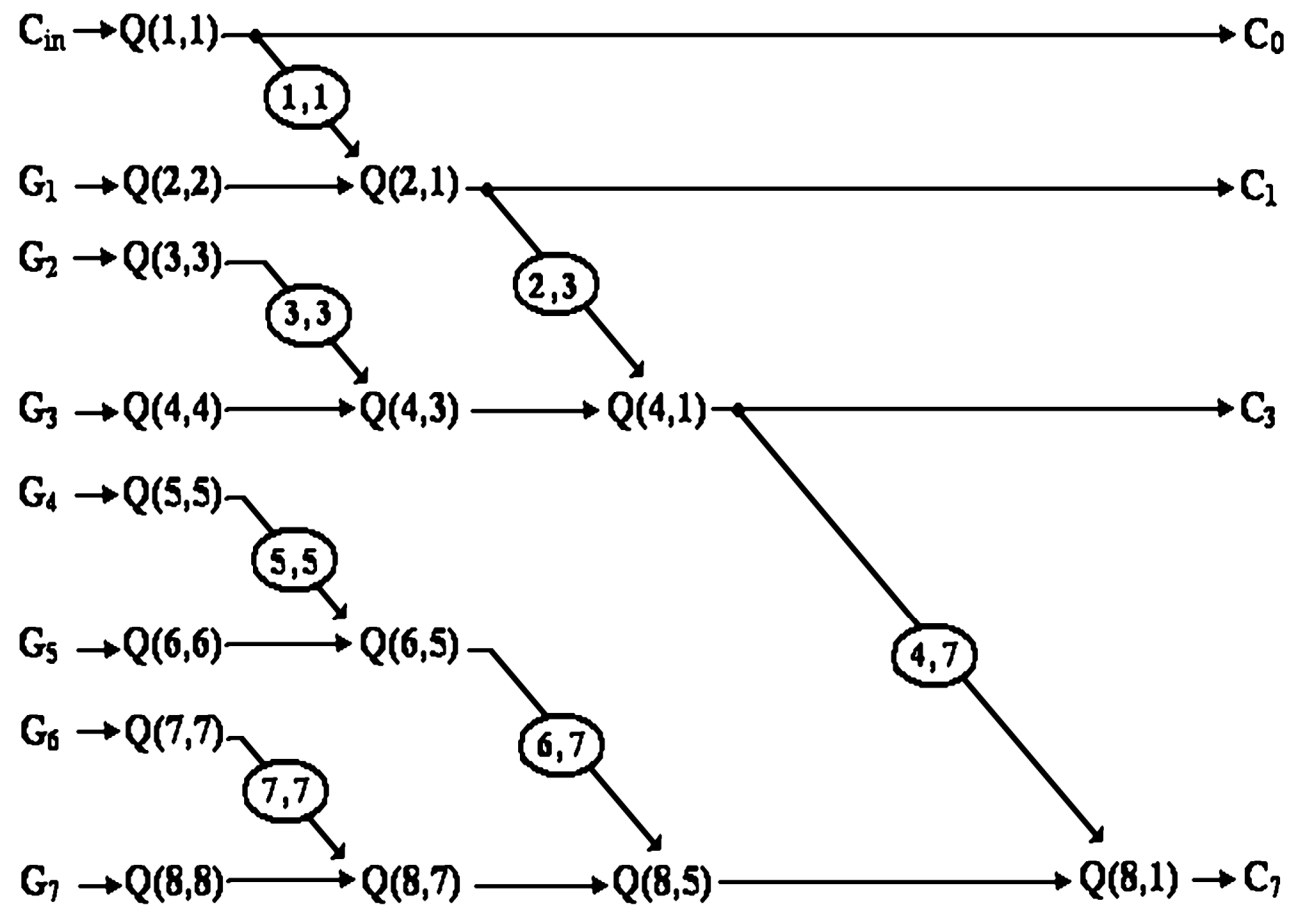

Figure 11. Sparsity-4 logarithmic stage carry-tree adder.

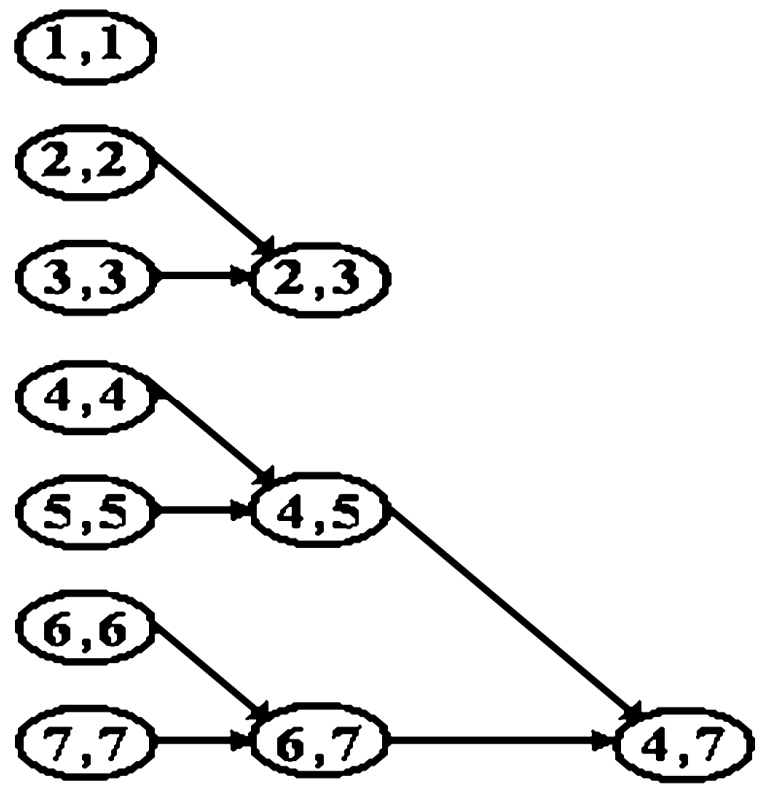

Figure 12. Product tree for sparsity-4 logarithmic stage carry-tree adder.

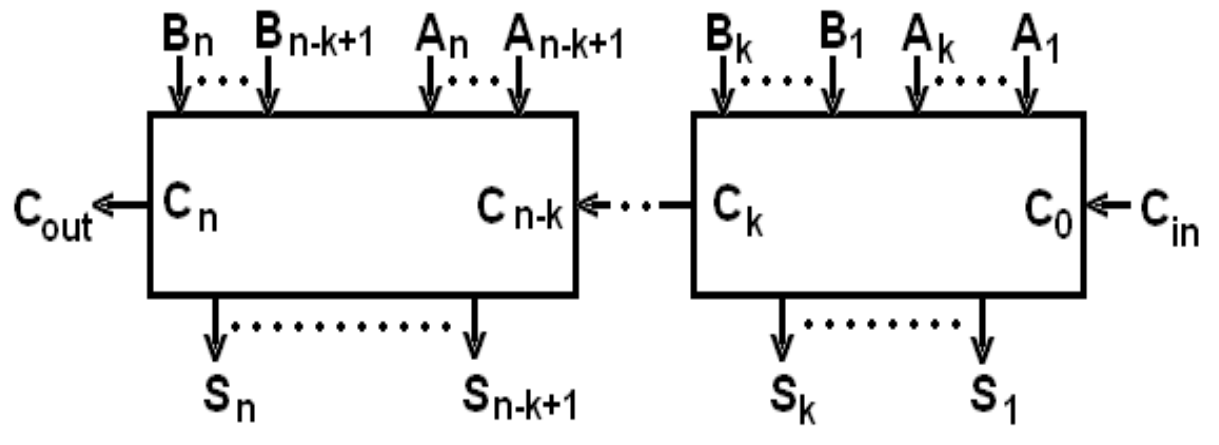

Figure 13. Block-ripple-carry optimization technique using in-block parallel adders.

## VII. Conclusion:

In this paper we have described a new scheme for quaternary logic which corresponds well with binary logic. This new scheme is used to design half adder and full adder in quaternary logic. We have proposed proficient designs of adder which are better than ripple carry adder having *n* gate-delay for *n* qudits. Single-stage carry look-ahead adder works very fast but has large fan-in for large number of qudits. So we have proposed logarithmic stage carry look-ahead adder which works within $\log_2(n)$ gate-delay for *n* qudits and have limited number of fan-in. Later we have analyzed the performance of different adders and have proposed a sparsity-4 logarithmic stage adder which merges the advantages of single-stage carry look-ahead adder and logarithmic stage carry-tree adder. We think there may be room for further improvement of the design of adder which can be really a promising field for future researchers. Extensive research in this sector will enhance the performance of processor as addition is the heart of computer arithmetic.

## References

[1] Brent, R. P.; Kung, H. T.; "A Regular Layout for Parallel Adders", *Computers*, *IEEE Transactions on*, Vol. 31, No. 3, March 1982, pp. 260-264.

[2] Kogge, P.; Stone, H.; "A Parallel Algorithm for the Efficient Solution of a General Class of Recurrence Equations", *Computers*, *IEEE Transactions on*, 1973, Vol. 22, No. 8, August 1973, pp. 783-791.

[3] Wei, B. W.; Thompson, C. D.; "Area-Time Optimal Adder Design", *Computers*, *IEEE Transactions on*, Vol. 39, No. 5, May 1990, pp. 666-675.

[4] Huntington, E. V.; "Sets of Independent Postulates for the Algebra of Logic", *Transactions of the American Mathematical Society*, Vol. 5, No. 3 , Jul. 1904, pp. 288-309.

[5] Jahangir, I.; Hasan, D. M. N.; Islam, S.; Siddique, N. A.; Hasan, M. M.; "Development of a novel quaternary algebra with the design of some useful logic blocks", *ICCIT 2009 - 12th International Conference on Computers and Information Technology*, pp. 197-202, Dhaka, 21-23 Dec. 2009.

[6] Jahangir, I.; Hasan, D. M. N.; Reza, M. S.; "Design of some quaternary combinational logic blocks using a new logic system", *TENCON 2009 - 2009 IEEE Region 10 Conference*, pp. 1 - 6, Singapore, 23-26 Jan. 2009.

[7] Awwal, A. A.; Munir, S. M.; Khalid, A. T. M. S.; Michel H. E.; Garcia, O.; "Multivalued Optical Parallel Computation Using an Optical Programmable Logic Array," *Informatica*, vol. 24, 2000, pp. 467-473.

[8] Current, K.W.; "A CMOS quaternary threshold logic full adder circuit with transparent latch", *ISMVL 1990 - 20th International Symposium on Multiple-Valued Logic*, pp. 168 - 173, Charlotte, 23-25 May 1990.

[9] Silva, R. C. G.; Boudinov, H. I.; Carro, L.; "A low power high performance CMOS voltage-mode quaternary full adder", *2006 IFIP International Conference on Very Large Scale Integration*, pp. 187 - 191, Nice, 16-18 Oct. 2006.

[10] Wu, X.; Chen, X.; Prosser, F.; "An investigation into quaternary CMOS full-adder based on transmission function theory", *ISMVL 1989 - 19th International Symposium on Multiple-Valued Logic*, pp. 58 - 62, Guangzhou, 29-31 May 1989.

[11] Khan, M. H. A.; "A recursive method for synthesizing quantum/reversible quaternary parallel adder/subtractor with look-ahead carry", *Journal of Systems Architecture: the EUROMICRO Journal*, Vol. 54, No. 12, Dec. 2008, pp: 1113-1121.